\documentclass{jpp}
\usepackage{color}
\usepackage{babel}
\usepackage{amsmath}
\usepackage{amssymb}
\usepackage{graphicx}
\usepackage[unicode=true,
 bookmarks=false,
 breaklinks=false,pdfborder={0 0 1},backref=section,colorlinks=false]
 {hyperref}

\makeatletter
\usepackage{epstopdf}
\usepackage{epsfig}

\usepackage{amsfonts}\usepackage{subfigure}
\usepackage{babel}

\makeatother

\begin{document}
\title{Plasma Rotation Induced by Biasing in Axially Symmetric Mirrors}
\author{A. D. Beklemishev \aff{1}\aff{2}\corresp{\email{beklemishev@inp.nsk.su}}}

\maketitle
\affiliation{\aff{1}Budker Institute of Nuclear Physics, Novosibirsk,
Russia \aff{2}Novosibirsk State University, Novosibirsk, Russia} 
\begin{abstract}
Physics of the plasma rotation driven by biasing in linear traps is
analyzed for two limiting cases. The first, relevant for traps with
decent plasma parameters, considers the line-tying effects to be responsible
for the drive as well as for the dissipation of the angular momentum.
Meanwhile, in long and thin traps with low plasma parameters or developed
turbulence the radial transport of the angular momentum becomes its
primary loss channel. The momentum flux goes into the scrape-off layer;
that makes conditions there partially responsible for the achievable
rotation limits. 
\end{abstract}
\keywords{open trap, plasma rotation, biasing}

\section*{Introduction}

Plasma rotation in crossed fields is present in most mirror traps,
it is both the naturally occurring phenomenon and a means to influence
the quality of confinement. Rotation is an important factor influencing
axial confinement (through the plasma potential), stability of discharges
and turbulent transport processes. In extreme cases (in centrifugal
traps) fast rotation can also alter the equilibrium of radial forces.
Rotation can be induced by many different factors, such as the radial
distribution of the ambipolar potential, biasing of plasma-facing
electrodes, injection of angular momentum with neutral beams, or the
asymmetric confinement of particles. In equilibrium the net sources
should be exactly balanced by sinks. The loss of the angular momentum
can also be due to multiple causes, the most prominent of these are
the line-tying and collisional or turbulent viscosity.

In this paper our aim is restricted to theoretical description of
plasma rotation induced by biasing in axially symmetric mirrors. The
work is stimulated by extensive experimental data on plasma flow and
rotation obtained on GDT \citep{Bagryansky_2015} and SMOLA \citep{Sudnikov_2022}
devices in the Budker Institute of Nuclear Physics (BINP). Recent
experiments on SMOLA (with relatively low plasma temperatures) demonstrated
radically different spatial profiles of rotation as compared to earlier
experiments on GDT. In particular, in contrast to the radial rotation
shear observed and used in the ``vortex confinement'' regimes of
GDT, the observed plasma rotation in SMOLA is radially rigid but sheared
axially. The axial shear, i.e. the gradient of rotation frequency
along the trap, is rather difficult to understand in circumstances
when the electron temperature is far below the applied biasing potential
and thus the electron pressure effects in Ohm's law should be negligible.

To outline the theoretical model and the problem under consideration
we should start with discussion of terms. The local rotation velocity
can be defined as the hydrodynamic mass-flow velocity. In the drift
approximation disregarding the electron inertia it can be decomposed
into the drift velocity of ions (including the diamagnetic drift and
the guiding center drift velocities) and the parallel flow rotation
(in presence of azimuthal magnetic field and parallel plasma flow).
Due to relatively small axial currents in linear traps in what follows
we can disregard the parallel flow rotation, while the drift velocity
of ions satisfies the hydrodynamic equation of ion motion. The two-fluid
description is insufficient for plasmas with intense neutral beam
heating such as in GDT. In such cases the two-fluid model should be
supplemented with an adequate description of the hot-ion component.

Equations describing the plasma dynamics should be coupled with corresponding
boundary conditions. The term ``biasing'' is referring to a specific
type of these. Magnetically confined plasma is typically surrounded
by various solid surfaces (usually conducting) such as limiters, end-plates,
vessel walls, etc. Different parts of the plasma-facing surfaces can
be grounded, floated or turned into biased electrodes by using external
power sources. However, the surface potential is not directly defining
the potential of the adjacent plasma due to the Debye shielding. In
dense plasmas of fusion traps the Debye radius is negligible, so that
biasing an electrode is meaningful only if there is a possibility
of its current connection to the plasma. Then, in a period of time,
it may lead to a change in the plasma potential.

As is known from the theory of plane probes \citep{Sheath} the spatial
structure of the electrostatic potential near the electrode is rather
complex: it contains a thin non-neutral ``sheath'' and a rather
wide quasi-neutral ``presheath'' layers. The potential jump in the
sheath layer depends on the electron temperature and the balance of
ion and electron currents to the electrode. The structure of the presheath
is determined by the distribution of the electron temperature and
density along the field lines and thus by the plasma-flow regime to
the electrode. In linear traps, in particular, the distributed potential
in expanders can be interpreted as a type of presheath. The parallel
electron current is the defining factor in the sheath potential, while
its influence on the presheath is negligible unless the electrode
is close to the ion-saturation regime. Thus it can be concluded that
the potential difference between the plasma core and the electrodes
(outside of saturation regimes) is composed of two parts: the full
ambipolar potential, depending on equilibrium parameters of the plasma
flow, and the difference between the real and the ambipolar sheath
potentials. The second part is affected by the electron current density
and thus has very little inertia.

Each open flux tube has two ends on solid surfaces, thus there are
two boundary conditions involving the plasma potential and the current
density at boundaries. Flux tubes can be longer or shorter than the
theoretical presheath. In the first case we can assume that there
is some `` core'' plasma between boundaries, while the short flux
tubes can typically be found between limiters and constitute the scrape-off-layers
(SOL). In the short flux tubes the plasma content is small since it
is due to the balance between cross-field particle transport and particle
sources within them, and fast parallel losses to the ends. Low plasma
density in SOL does not mean these discharge areas are unimportant
or equivalent to vacuum. They are providing the edge boundary conditions
for the core of the plasma column.

To induce the plasma rotation by biasing it is necessary to propagate
the transverse electric fields between the edge electrodes along the
field lines into the core plasma. Then the plasma rotation will change
due to the resulting radial electric field in it. The transition process
is equivalent to an Alfven wave of twisting propagating from the edge.
For transitions slower than the Alfven transit time the field-line
twisting can be ignored and the process can be described as electrostatic.
Electrostatic and electromagnetic cases differ only in whether there
is a parallel inductive electric field along the field lines. In any
case the angular acceleration of the plasma is due to the Ampere's
force on the radial currents (mainly in ions) which are closed by
the parallel electron currents from the electrodes. Alternatively
it can be said that any angular acceleration causes charge-separating
radial drift of plasma particles that is neutralized by axial electron
currents to the electrodes.

The plasma currents related to its accelerated flow in the magnetic
field, especially the fast outflow of rotating plasma in expanders,
can alter the weak magnetic fields there. This problem is obviously
complex and requires self-consistent solution. It is addressed in
the recent work by \citet{smolyakov2023alfvenization}. Here we will
assume the magnetic field perturbations to be small or adequately
described according to the cited work.

The plasma inertia requires finite charge to flow from the power source
through the sheaths in order to change the rotation velocity. This
process is extended in time and accompanied by energy dissipation
at the electrodes. In terms of electric circuits the sheath structures
are similar to thermal voltage sources with nonlinear internal resistance
while the rotating plasma core is equivalent to a capacitor.

Large-area metal endplates were used in early designs of mirrors to
provide the \textit{line-tying} stabilization. The idea is that at
low electron temperature (or with unlimited electron emission) the
sheath potential and its resistance can be neglected and then the
plasma potential also becomes uniform. The ExB plasma drifts are suppressed.
At non-zero temperatures such electrodes project non-zero radial electric
fields into the plasma core due to radial dependence of the sheath
potential and lead to ambipolar plasma rotation. This may become a
cause for the temperature-gradient instability of flute modes \citep{10.1063/1.859699}.
The sheath resistivity also becomes finite, so that the line-tying
just slows the growth rates of flute modes. For modes growing faster
than the circuit recharge time the line-tying fails completely. The
sheath resistivity can be reduced by higher emissivity of electrodes
but such setup effectively limits the electron temperature to a few
volts.

The GDT group uses advanced type of flute-mode suppression called
the `` vortex confinement'' \citep{Beklemishev2010VortexCO}. The
end plates in this case look like relatively biased concentric rings.
Alternatively, the biasing voltage can be applied between the central
end-plates and the limiters. Such arrangement gives rise to a layer
of sheared plasma rotation around the flux surface corresponding to
the boundary between biased plates. Fast mode rotation plus the sheath
dissipation (as in line-tying) lead to nonlinear dissipative saturation
of unstable flute modes, somewhat similar to resistive wall stabilization
of edge modes in tokamaks. Although both signs of the radial electric
field, and thus both rotation directions, are suitable, better result
can be achieved with negative biasing potential on axis (or with positive
potential on limiters). In this case the induced ExB rotation in GDT
discharges is opposite to the ambipolar rotation and the diamagnetic
drift, thus effectively reducing the centrifugal instability drive.
Furthermore, the net radial current, driven by the external voltage
source, is in ions, and is directed inward, leading to an additional
pinch effect.

The discharge in the SMOLA device \citep{Sudnikov_2022} is driven
by the plasma gun with heated LaB cathode that covers most of the
discharge crosssection. The end-plate opposite the cathode is composed
of concentric Mo ring electrodes, biased approximately to the cathode
potential. The outer concentric anode and limiters can be biased or
grounded, while the voltage between the cathode and the outer electrodes
can serve to drive the plasma rotation. This setup is rather generic
for production of rotating plasma discharges. In SMOLA the high rotation
frequency is one of the objectives of the experiment, so that it is
measured in detail by spectroscopic means and by Mach probes directly,
and via electric field probes indirectly. As mentioned above, the
observed results are quite different from those initially expected.
Even the scaling of the rotation frequency with the magnetic field
is somewhat off. In this paper we argue that the possible causes of
the anomalous structure of rotation are several types of dissipation
prominent in SMOLA experiments.

The first section of this paper presents essential equations for stationary
rotating discharges. In the second section the physical model is simplified
for description of plasma rotation driven by biasing in GDT-type traps.
The third section describes the alternative model, relevant for plasmas
with high effective viscosity (or discharges with small relative radii).
This model emphasizes the role of plasma-SOL interaction and is intended
for interpretation of SMOLA experiments. In Conclusion the results
are summarized.

\section{Standard model of plasma rotation in GDT}

Consider the quasi-stationary state of the axially symmetric plasma
discharge (on a time scale larger than the Alfven time). In this case
the electric field can be considered electrostatic and described by
means of the local plasma potential distributed on a flux surface
$\psi$ along each field line, $\phi=\phi(\psi,z)$. Furthermore,
due to high electron temperature in this model we will neglect the
parallel resistivity of the discharge. Then the parallel electric
field within the quasineutral bulk plasma is due entirely to the gradient
of the electron pressure. Within the core plasma the electric field
ensures quasineutrality via the Boltzmann distribution of electrons.
In the expanders (presheath) the distribution of electrons deviates
from the equilibrium but the potential profiles along a field line,
including the non-neutral sheaths, can be calculated \citep{Skovorodin2019,10.1063/5.0171845}.

\begin{figure}
\includegraphics[width=1\columnwidth]{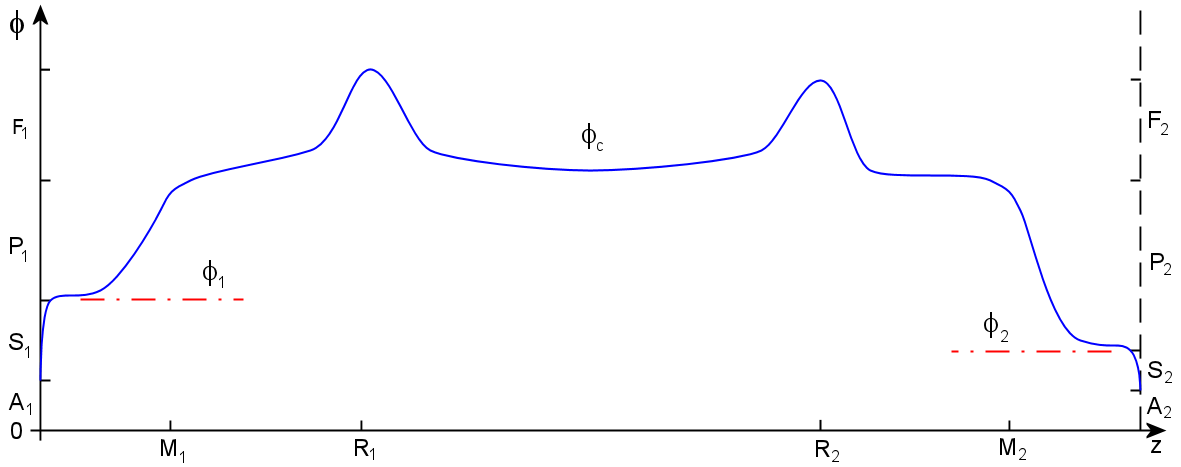}

\caption{Scheme of the potential distribution along a field line in GDT-type
linear traps. Subscripts 1,2 denote values relative to west and east
ends of the device, $M_{i}$ and $R_{i}$ are the axial positions
of mirror throats and reflection points for sloshing hot ions, respectively.
$A_{i}$ are the voltages applied to the end-plates, $S_{i}$ and
$P_{i}$ are the sheath and presheath (expander) potentials, $F_{i}$
are the heights of the ambipolar barriers due to the density peaks
of hot ions at the reflection points. \label{fig:Scheme1}}
\end{figure}

The scheme of the potential distribution along internal field lines
in GDT-type linear traps is presented in Fig.1. The potential profile
in the quasineutral region, including the core plasma and the expander
volume, is essentially determined by the electron temperature and
the density distribution, i.e. by the regime of the axial plasma confinement
and outflow. As a result, the potential difference $\phi_{1}-\phi_{2}$
is a function of said parameters too. Meanwhile, the sheath potential
difference $\Delta S=S_{1}-S_{2}$ depends on the parallel current
densities to the end-plates, $j_{1},j_{2}$. Thus the sheath potentials
balance the external ($A_{1}-A_{2}$) and the internal ($\phi_{1}-\phi_{2}$)
voltages in the presence of the electron currents, making the plasma
flow in expanders in general non-ambipolar.

The two plasma currents $j_{1}(S_{1})$ and $j_{2}(S_{2})$ have opposite
directions due to orientation of end-plates, but let's set their signs
the same (as measured by external circuits). Such convention is suitable
for relating the currents to the volt-ampere characteristics of sheaths
(ignoring their orientation). It is also convenient to measure the
parallel current densities per unit flux, i.e. $j_{i}\equiv\partial I_{i}/\partial\psi\propto j_{\parallel}/B.$
Then $j_{1},j_{2}$ can be rewritten in terms of the `` net'' current
through the trap $j_{n}=(j_{2}-j_{1})/2$ and the `` biasing'' current
$j_{b}=j_{1}+j_{2}$ (its closure is by the radial currents). For
small currents, in the linear approximation, $\Delta S$ will depend
only on the net current $j_{n}$, but in experiments the logic is
reversed, $j_{n}$ is a function of $\Delta S$. For example, the
difference in gas conditions and expansion ratio in the east and west
ends of the GDT (due to positions of gas puffing and plasma gun placement)
results in a significant net axial current (of the order of the ion
current) on field lines with $A_{1}=A_{2}$. The net current can be
used for `` direct energy conversion'' , but it results in enhanced
energy losses of plasma electrons. Alternatively, the best axial electron
confinement can be achieved for the minimum net axial current. This
requires application of a suitable external voltage $A_{1}-A_{2}$
between opposite end-plates. This is a feedback-type problem since
such external voltage depends on plasma conditions and varies in time
during the discharge. Alternatively, the end-plates can be isolated,
but in this case the ability to set the biasing potential $\phi_{b}=(A_{1}+A_{2})/2$
is lost. A rough type of balancing can be achieved by setting a diode
circuit on one and the voltage source on the opposite end-plates.
In the following we will assume that some kind of balancing is in
place, so that the current density to the end plate is less than the
ion loss rate, while the functional dependence $j_{i}(S_{i})$ is
approximately linear.

In order to relate Fig.\ref{fig:Scheme1} to the plasma rotation we
need to consider dependence of shown potentials on the radial position
of the field line, i.e. their dependence on the flux coordinate. One
part of this functional dependence stems from the radial distribution
of electron temperature and plasma density within the discharge, the
ambipolar potential$\phi_{a}(\psi,z)$, the other is due to the radial
dependence of the applied biasing voltage on the end-plates, $\phi_{b}(\psi)$,
while the third reflects the state of the angular momentum transfer
due to radial currents, $\phi_{m}(\psi)$. The sheath potential $S_{i}$
contains contributions of the ambipolar and non-ambipolar nature,
$S_{i}=S_{ia}+S_{in}$. This means that the sheath potential calculated
with the assumption of equal electron and ion loss rates, i.e. $j_{i}=0$,
is $S_{ia}$. At small currents $S_{i}\approx S_{ia}+j_{i}(\partial S_{i}/\partial j_{i})_{a}$,
$S_{in}\approx j_{i}(\partial S_{i}/\partial j_{i})_{a}$. However,
both $S_{ia},S_{in}$ result in $z$-independent contributions to
the overall potential, so that the ambipolar part of the sheath potential
can be included either in $\phi_{a}$ or in $\phi_{m}$. For the sake
of brevity in terms we will assume that $\phi_{m}$ is non-ambipolar,
$\phi_{m}=(S_{1n}+S_{2n})/2\propto j_{b}$, while $\phi_{a}$ corresponds
to the idealized axial equilibrium with $j_{b}=0$ and zero radial
currents.

In summary, the plasma potential distribution can be written as a
sum of the ambipolar potential, $\phi_{a}$, the biasing potential,
$\phi_{b}$, and the non-ambipolar sheath potential, $\phi_{m}$ :
\[
\phi(\psi,z)=\phi_{a}(\psi,z)+\phi_{b}(\psi)+\phi_{m}(j_{b}(\psi)).
\]
In the paraxial approximation $d\psi\approx2\pi rBdr$, so that the
ExB drift velocity can be described as 
\[
v_{E}=c\frac{-\phi'_{r}}{B}\approx2\pi rc\frac{\partial\phi}{\partial\psi}.
\]

The plasma rotation velocity can be found from the radial component
of the equation of motion for the ion fluid (here we again assume
a paraxial magnetic structure): 
\[
-nM\frac{v^{2}}{r}=-p'_{r}-qn\varphi'_{r}+\frac{qn}{c}vB,
\]
where $v$ is the azimuthal velocity, $p'_{r}$ is the radial pressure
gradient, $q$ is the ion charge and other notations are conventional.
As a result, 
\[
\omega=\frac{v}{r}=\frac{\Omega}{2}\left(\sqrt{1+4\frac{v_{E}}{\Omega r}+\frac{4p'_{r}}{nM\Omega^{2}r}}-1\right),
\]
where $\Omega=qB/Mc$ is the ion cyclotron frequency. In the limit
of slow rotation, $2\omega/\Omega\ll1$, when the centrifugal effect
is negligible, 
\[
\omega\approx v_{E}/r+p'_{r}/(nM\Omega r)\equiv\omega_{E}+\omega_{d},
\]
where $\omega_{d}(\psi,z)=p'_{r}/(nM\Omega r)$ is the diamagnetic
drift frequency of ions. In this limit 
\[
\omega(\psi,z)=\omega_{d}(\psi,z)+2\pi c\frac{\partial\phi}{\partial\psi}\approx\omega_{da}(\psi,z)+\omega_{bm}(\psi).
\]
Here $\omega_{da}(\psi,z)=\omega_{d}(\psi,z)+2\pi c(\partial/\partial\psi)\phi_{a}$
is fully determined by the equilibrium plasma parameters, while $\omega_{bm}(\psi)=2\pi c(\partial/\partial\psi)(\phi_{b}+\phi_{m})$
is related to biasing and the conservation of the angular momentum.
The important point here is that the rotation frequency$\omega_{bm}(\psi)$
is `` rigid'' along the field lines (in the presheath and the plasma
core), i.e. is constant on each flux surface. In particular, formally
increasing the biasing potential far above the plasma temperature,
$\omega_{da}(\psi,z)\ll\omega_{bm}(\psi)$, we should get a rigid
rotation (along the field lines).

As mentioned above, $\phi_{m}$ and $\omega_{bm}$ are related to
the radial currents, which, in turn, affect the angular velocity.
These relationships can be described by the azimuthal component of
the fluid equation for the plasma as a whole coupled with the current
closure condition on a flux surface: 
\[
nM\frac{dv}{dt}=-\frac{1}{c}j_{\psi}B+f,
\]
\[
\left(\vec{b}\nabla\right)\frac{j_{\parallel}}{B}+\frac{\partial}{\partial\psi}\left(2\pi rj_{\psi}\right)=0.
\]
Here $j_{\psi}$ is the transverse (radial) current density, $f$
represents the azimuthal non-electromagnetic force per unit volume.
In particular, $f$ can stand for the divergence of the viscous momentum
flux, or for the collisional momentum transfer from the beam ions
or by charge exchange with neutrals. For simplicity we again used
the paraxial approximation

The radial current density is very difficult to measure experimentally
and thus it should be excluded from the system. We find 
\[
\frac{\partial}{\partial\psi}\left(2\pi rj_{\psi}\right)=\frac{\partial}{\partial\psi}\frac{2\pi rc}{B}\left(f-nM\frac{dv}{dt}\right)=-\left(\vec{b}\nabla\right)\frac{j_{\parallel}}{B},
\]
\begin{equation}
\frac{\partial}{\partial\psi}\left(\frac{rnM}{B}\frac{d}{dt}r\omega\right)=\frac{1}{2\pi c}\left(\vec{b}\nabla\right)\frac{j_{\parallel}}{B}+\frac{\partial}{\partial\psi}\left(\frac{rf}{B}\right).\label{eq:dodt}
\end{equation}
This equation describes the temporal evolution of the rotation frequency,
$\omega$. However, as we found earlier, this frequency is comprised
of two components $\omega=\omega_{da}(\psi,z)+\omega_{bm}(\psi),$of
which the first one is essentially an external parameter, i.e. defined
by equations of mass and energy transport outside of this model. Meanwhile,
the variable part of the frequency $\omega_{bm}(\psi,t)$ is `` rigid'' ,
so that the equation for it can be reduced to the one-dimensional
form by averaging along the field lines. After integrating Eq.(\ref{eq:dodt})
along field lines between the opposite end-plates we get: 
\begin{equation}
\frac{\partial}{\partial\psi}\left(\left\langle \frac{r^{2}nM}{B}\right\rangle \dot{\omega}_{bm}+U\omega_{bm}\right)=\frac{j_{b}}{2\pi c}+\frac{\partial}{\partial\psi}\left\langle \frac{r}{B}\left(f-nM\frac{d}{dt}r\omega_{da}\right)\right\rangle .\label{eq:dodt1}
\end{equation}
Here we took into account that 
\[
\int\left(\vec{b}\nabla\right)\frac{j_{\parallel}}{B}d\ell=\left.\frac{j_{\parallel}}{B}\right|_{z_{1}}^{z_{2}}=j_{2}+j_{1}=j_{b},
\]
and introduced notations 
\[
\left\langle \right\rangle =\int_{z_{1}}^{z_{2}}d\ell,\qquad U(\psi)=\left\langle \frac{rnM}{B}v_{\parallel}\left(\vec{b}\nabla\right)r\right\rangle .
\]
Coefficient $U$ accounts for the angular momentum transfer with the
axial plasma flow. Strictly speaking this is a non-paraxial effect,
but it can be important in case of intense flows in expanders. Namely,
taking into account the rigid rotation frequency, the loss rate of
the angular momentum with the outflow in expanders will be larger
due to the increasing radius of rotation at the end-plates.

Equation (\ref{eq:dodt1}) can describe the evolution of $\omega_{bm}(\psi,t)$
if its coefficients and the source terms in the right-hand side are
known. However, the line-tying current density $j_{B}$ depends on
the non-ambipolar sheath potential $\phi_{m}$, which contributes
to $\omega_{bm}$ itself: 
\[
\omega_{bm}=2\pi c\frac{\partial}{\partial\psi}\left(\phi_{b}+\phi_{m}\right),
\]
\begin{equation}
j_{b}=j_{b}(\phi_{m})=j_{b}\left(\frac{1}{2\pi c}\int\omega_{bm}d\psi-\phi_{b}\right).\label{eq:IV}
\end{equation}
Explicit form of relation (\ref{eq:IV}) depends on the IV characteristic
of the sheath layer. For low-emission endplates and low net axial
currents (below the ion saturation threshold) relation (\ref{eq:IV})
can be linearized and rewritten as 
\begin{equation}
j_{b}=j_{ion}\frac{e\phi_{m}}{T_{e}}=j_{ion}\frac{e}{T_{e}}\left(\frac{1}{2\pi c}\int\omega_{bm}d\psi-\phi_{b}\right),\label{eq:jb}
\end{equation}
where $j_{ion}=J_{i}/B$ is the ion current density (per flux tube)
to the end-plate. In particular, such model was used earlier in {[}vortex{]}.

There is at least one more term in the right-hand side of (\ref{eq:dodt1})
that depends on $\omega_{bm}$. It is the corresponding viscous force
\begin{equation}
f_{\eta b}=\frac{\eta}{r^{2}}\frac{\partial}{\partial r}\left(r^{3}\frac{\partial\omega_{bm}}{\partial r}\right),\label{eq:eta}
\end{equation}
where $\eta$ is the hydrodynamic viscosity. In fusion plasmas it
is usually considered to be very small but can be anomalous in presence
of turbulence. This term contains the highest radial derivative.

Taking into account the above relations, Eq.(\ref{eq:dodt1}) can
be rewritten in terms of the plasma potential induced by biasing,
$\phi=\phi_{b}+\phi_{m}$: 
\begin{equation}
\frac{\partial}{\partial\psi}\left(I\frac{\partial\dot{\phi}}{\partial\psi}+U\frac{\partial\phi}{\partial\psi}\right)+H\left(\phi-\phi_{b}(\psi)\right)-\frac{\partial^{2}}{\partial\psi^{2}}\left(\Upsilon\frac{\partial^{2}\phi}{\partial\psi^{2}}\right)=F,\label{eq:fullphi}
\end{equation}
where 
\[
I=\left\langle \frac{r^{2}nM}{B}\right\rangle ,\qquad F=\frac{\partial}{\partial\psi}\left\langle \frac{r}{2\pi cB}\left(\hat{f}-nM\frac{d}{dt}r\omega_{da}\right)\right\rangle ,
\]
\[
\Upsilon=4\pi^{2}\left\langle \eta Br^{4}\right\rangle ,\qquad H=j_{ion}\frac{e}{4\pi^{2}c^{2}T_{e}},
\]
$H$ corresponds to the sheath-adjacent plasmas and angular brackets
denote integration along the field lines. Note that the biasing potential
$\phi_{b}$ enters through the line-tying term $H\left(\phi-\phi_{b}(\psi)\right)$
only.

\begin{figure}
\includegraphics[width=0.3\textwidth]{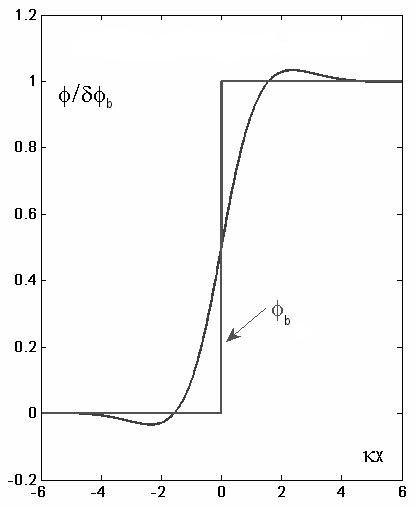}\caption{\label{fig:Equilibrium-profile}Equilibrium profile of the rotation
potential driven by biasing adjacent endplates and regularized by
viscosity. Here $x=\psi-\psi_{0}$, $\kappa=\left(4H/\Upsilon\right)^{1/4}.$ }
\end{figure}

As an example let us consider rotation of the plasma column with radially
uniform density and electron temperature, driven by biasing adjacent
concentric rings of end-plates (as in the `` vortex confinement''
\citep{Beklemishev2010VortexCO}.) Then $\phi_{b}(\psi)$ is discontinuous
like a Heaviside function at $\psi=\psi_{0}$. Further assume the
plasma to be in equilibrium with low axial losses and no momentum
injection with beams, $F=0,U=0.$ Then Eq.(\ref{eq:fullphi}) describes
how the rotation potential $\phi$ tries to relax to the applied biasing
potential $\phi_{b}$ but is hindered by viscosity. If $\eta$ is
small, the rotation layer in equilibrium is in a small area around
$\psi_{0}$, and the solution can be found analytically. It was earlier
published in \citep{Beklemishev2010VortexCO} and is shown in Fig.\ref{fig:Equilibrium-profile}.
The main point here is the weak dependence of the solution width $\Delta$
around $\psi_{0}$ on viscosity 
\[
\Delta=(\Upsilon/4H)^{1/4}.
\]
In GDT the solution width is of the order of 20\% of the plasma radius
(by estimates and by probe measurements). If this width were comparable
to the discharge radius, we would need to solve the boundary problem
and study the boundary conditions for Eq.(\ref{eq:fullphi}) (in $\psi$)
in detail.

In summary, frequency of the plasma rotation in GDT-like devices can
be described as having two distinct components: 1) the `` ambipolar''
frequency $\omega_{da}(\psi,z)$, which is a function of distributions
of density and temperature in the discharge, and 2) the `` driven''
frequency $\omega_{bm}$, which reflects the radial distribution of
the angular momentum. If the axial ion confinement is independent
of the `` driven'' rotation, then $\omega_{bm}$ is constant on
each flux surface and its temporal evolution can be found from the
equation of the angular momentum transport, (\ref{eq:fullphi}).

\section{Rotation of low-temperature plasmas}

\begin{figure}
\includegraphics[width=0.6\textwidth]{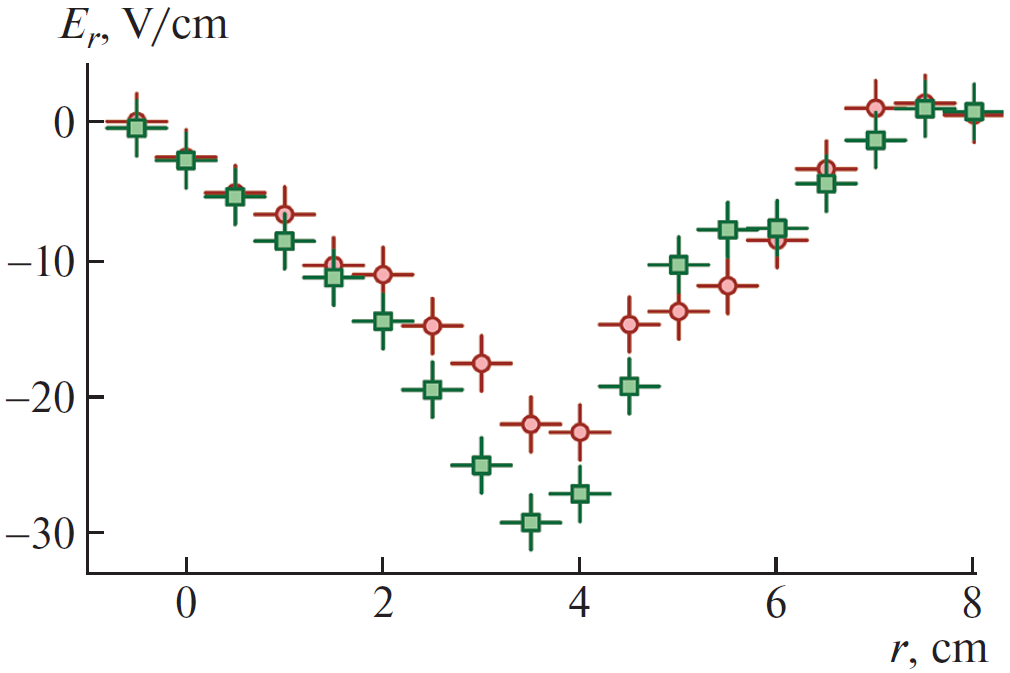}\caption{\label{fig:Radial-distribution-of}Radial distribution of the radial
electric field in SMOLA at Z = 2.04 m in different magnetic field
configurations \citep{Inzhevatkina2024}. Approximately linear profile
in the plasma core ($r<3.8$ cm) corresponds to rigid rotation, while
in the SOL ($r>3.8$ cm) the rotation velocity decays to zero.}
\end{figure}

Plasma rotation in the SMOLA device \citep{Sudnikov_2022} exhibits
rigid-rotor type of behavior in radius up to the SOL boundary for
all types of the biasing distribution. It can be seen from the recently
published probe measurements of the radial electric field \citep{Inzhevatkina2024}
shown in Fig.\ref{fig:Radial-distribution-of}.

The radially rigid rotation indicates that, unlike in GDT, the effective
viscous transport of momentum in radius is dominant. It should be
noted that in these experiments the electron temperature is less than
10\% of the biasing potential, while the diamagnetic drift frequency
should also be small. This means that in terms of the previous section
the ambipolar frequency $\omega_{da}$ can be neglected. There are
also no obvious non-electromagnetic sources of the angular momentum
within the discharge core, $\hat{f}=0.$

As compared to the above GDT-relevant model theoretical description
of SMOLA discharges should include effects of finite conductivity
and fast axial plasma flow. Indeed, the rotation and its potential
in SMOLA vary along the discharge. In conjunction with low electron
temperature it means that the parallel gradient of the plasma potential
is ohmic. Besides, the plasma gun of SMOLA generates a constant sub-sonic
axial flow of plasma through the device. This flow is sure to play
an important role in the axial transport of the angular momentum.

Let us construct a simplified model relevant for low-temperature viscous
plasmas. Assume that a cylindrical plasma column flows along the uniform
magnetic field $B$ and rotates in crossed fields. The ambipolar potential
is zero. Assume the rotation to be rigid in radius with frequency
$\varOmega(z)$ up to the ``edge'' at $r=a$ (in SOL, at $r>a$,
the rotation is no longer rigid.) Assumption of the rigid rotor replaces
equation of the momentum transport in radius, while the feature of
the model will be construction of equation for the axial transport
of the angular momentum.

For a rigid $E\times B$ rotation the plasma potential is parabolic
($v(r)\propto r$ and the edge potential is $\phi(a)=\varphi_{*}(z)$):
\[
\phi(r,z)=\frac{B}{2c}\varOmega(z)\left(r^{2}-a^{2}\right)+\varphi_{*}(z).
\]
The axial plasma current density can then be found from the Ohm's
law: 
\begin{equation}
j_{z}(r,z)=\sigma E_{z}=\frac{\sigma B}{2c}\varOmega'(z)\left(a^{2}-r^{2}\right)-\sigma\varphi_{*}'(z).\label{eq:jz1-1}
\end{equation}
The radial current density can be found from the current closure condition.
In case of radially uniform conductivity $\sigma(r)=const$ it is
also parabolic: 
\[
j_{r}(r,z)=-\frac{1}{r}\int_{0}^{r}\frac{\partial j_{z}}{\partial z}rdr=\sigma\frac{Br}{8c}\varOmega''(z)\left(r^{2}-2a^{2}\right)+\sigma\frac{r}{2}\varphi_{*}''(z).
\]

In our model there are two possible sources of torque: the Ampere's
force, $j_{r}B_{z}$, and the edge friction. Then the balance of the
angular momentum looks like 
\[
\frac{\partial\varPhi(z)}{\partial z}=M_{A}+M_{\eta}.
\]
Here $\varPhi$ is the axial flux of the angular momentum, 
\[
\varPhi(z)=2\pi\varOmega(z)\int_{0}^{a}\rho v_{z}r^{3}dr\equiv G\varOmega,
\]

\[
M_{A}=-\frac{2\pi}{c}\int_{0}^{a}j_{r}Br^{2}dr=\frac{\pi\sigma B^{2}}{12c^{2}}a^{6}\varOmega''(z)-\frac{\pi\sigma B}{4c}a^{4}\varphi_{*}''(z)
\]
is the torque of the Ampere's force, while the torque of the edge
friction, $M_{\eta}$, should be found from the boundary conditions
in radius.

\begin{figure}
\includegraphics[width=0.8\textwidth]{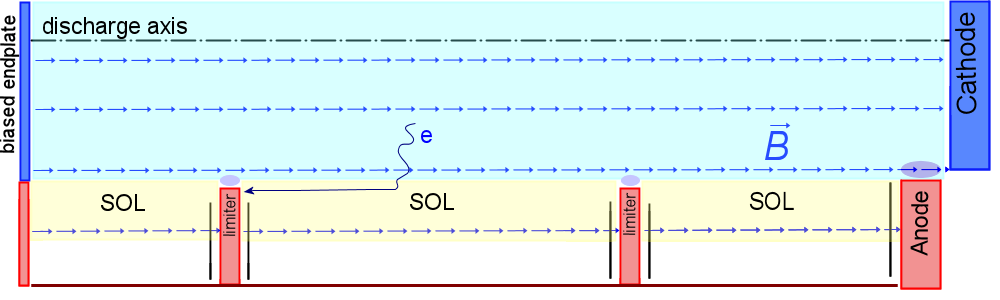}

\caption{\label{fig:Boundaries}Scheme of the structure and the boundary conditions
of the model discharge.}
\end{figure}

The structure of the model discharge as well as the realistic boundaries
are shown in Fig.\ref{fig:Boundaries}. Along the magnetic field $\vec{B}$
the discharge is in electrical contact with the cathode and the biased
endplate. There is a plasma source at the cathode and a sink at the
endplate. At $r=a$ the discharge is in contact with SOL (scrape-off
layer) plasma. The SOL plasma is in contact with the anode, limiters
and floating shielding. Some current may flow to the inner faces of
anode and limiters, or to their sides across SOL.

In general, areas on top of limiters are rather small and should not
exert too much of an influence on plasma dynamics (though this may
not be the case for the anode). These areas should exhibit the sheath
property of shielding the external electric potential of nearby electrodes
by higher density of electrons. However, the radial current density
there should be in electrons mostly (in absence of surface ion sources)
and thus be small, unless there is a significant azimuthal friction
force on electrons. Such force can be due to the neutral gas atoms
or the plasma ions. Both these contributing forces should be present
on the SOL boundary away from the limiters as well. The SOL boundary
has much larger area than the limiter tops, so that the radial current
flows mostly through this boundary and then to the sides of the limiters.
The anode top may be different due to higher gas density there and
its higher area in some trap designs.

Discussing the role of limiters it should also be noted that their
side surfaces act as electrodes with line-tying properties, i.e. provide
an effective friction for the plasma rotation due to the sheath dissipation.
This friction in principle can be mitigated by splitting the side
surfaces into radially isolated rings. Biasing the limiters should
be effective only if the plasma density in the SOL is sufficiently
high, since the biasing current is limited by the ion current from
the SOL to the limiter side. Shielding the sides of a limiter by floating
electrodes reduces the biasing current limit and ultimately becomes
equivalent to setting the whole limiter at the floating potential.

The SOL plasma experiences azimuthal radially differential flow, while
the axial flow can be neglected. Then, in the slab approximation (the
SOL width is usually smaller than its radius):

\[
\frac{\partial\varPi_{xy}}{\partial x}=-\frac{1}{c}j_{x}B-qv_{y},\qquad\varPi_{xy}=\rho v_{x}v_{y}-\eta\frac{\partial v_{y}}{\partial x},
\]
where $\varPi_{xy}$ is the radial flux of the azimuthal momentum,
$qv_{y}$ is the momentum transfer to neutrals. Thus, the radial flux
of momentum in SOL can decay either due to the Ampere's force via
currents to the limiters, or by recycling.

Using again the current closure and the Ohm's law, 
\[
\frac{\partial j_{x}}{\partial x}=-\frac{\partial j_{z}}{\partial z}=\sigma_{s}\frac{\partial^{2}\phi}{\partial z^{2}},
\]
in terms of the electrostatic potential we get ($\rho v_{x}\rightarrow0$,
$v_{y}=c/B(\partial\varphi/\partial x)$): 
\begin{equation}
\left(\frac{B^{2}\sigma_{s}}{c^{2}}\right)\frac{\partial^{2}\phi}{\partial z^{2}}=\eta\frac{\partial^{4}\phi}{\partial x^{4}}-q\frac{\partial^{2}\phi}{\partial x^{2}}.\label{eq:sol}
\end{equation}
Though equation (\ref{eq:sol}) is comprehensive, it is too complicated
to use, especially since the plasma parameters in SOL are usually
known rather poorly.

Consider the line-tying regime in SOL: \textcolor{black}{neglect recycling
($q=0$), integrate between limiters ($L_{m}$ is the line length),
and apply boundary conditions on them ($j_{z}=\pm J_{lim}$):} 
\[
\frac{\partial^{4}\phi}{\partial x^{4}}=\frac{2B^{2}J_{lim}}{\eta c^{2}L_{m}}.
\]
Assuming that the limiters are biased to a fixed potential $\phi_{lim}$
and the sheath current density is far from saturation, the current
density to the limiters is 
\[
J_{lim}\approx J_{i}(1-\exp(e(\phi_{lim}-\phi)/T_{e}))\approx eJ_{i}/T_{e}(\phi-\phi_{lim}),
\]
where $J_{i}$ is the ion flux density. Then 
\begin{equation}
d_{\eta}^{4}\phi^{''''}+\phi-\phi_{lim}=0,\label{deta-1}
\end{equation}
with the characteristic radial SOL width 
\[
d_{\eta}=\left(\frac{\eta c^{2}L_{m}T_{e}}{2J_{i}eB^{2}}\right)^{1/4}.
\]
This width can be estimated for SMOLA parameters as $d_{\eta}\sim0.6$
cm (with the classic viscosity $\eta$ \citep{1965B}). However, this
estimate is not really reliable and can be regarded as a rough lower
limit. In presence of turbulence $\eta$ and $d_{\eta}$ should be
larger, while the turbulence is in fact observed at the SOL boundary.
Another possible cause for the increase in $d_{\eta}$ is the saturation
of the electron sheath current. It occurs due to the fact that the
transverse current is mostly in ions, while the non-ambipolar current
to the limiters is in electrons. Thus the number of electrons that
can be extracted from each flux tube is limited. In other terms, the
presheath length can become larger than $L_{m}$. In this case the
line-tying effect weakens and the SOL width is enhanced.

Another possible plasma behavior in SOL can be denoted as the recycling
regime: neglect the Ampere's force (the left-hand side) in Eq.\ref{eq:sol},
then it yields: 
\[
\phi'''-\frac{q}{\eta}\phi'=0,\qquad\phi=\left(\varphi_{*}-\varphi_{0}\right)\exp\left(-x/d_{q}\right)+\varphi_{0},
\]
where $d_{q}=\sqrt{\eta/q}$, $\varphi_{0}$ is the potential of the
outer shell. For SMOLA $d_{q}\sim0.3-3$ cm, depending on estimates
of viscosity and the density of neutrals, that is comparable to $d_{\eta}$.
This means that both the recycling and the line-tying can play a role.

The matching conditions for the plasma core to the boundary of SOL
are as follows: the potential $\phi$ is continuous, 
\[
\varOmega=\frac{v_{y}(a)}{a}=\frac{c}{aB}\left.\frac{\partial\phi}{\partial x}\right|_{a},
\]
\[
M_{\nu}=2\pi a^{2}\eta\left.\frac{\partial v_{y}}{\partial x}\right|_{a},
\]
and $j_{\perp}(a)=j_{x}$. The last condition (of the current continuity)
is relevant for the line-tying regimes in SOL, while in the recycling
regime the current density plays no role in the plasma dynamics.

It is very difficult to measure the transport coefficients in SOL
in order to evaluate the matching conditions. However, the problem
can be reversed: we can obtain information about conditions in the
SOL by measuring the radial profile of potential there. Assume that
we can measure the gradient length $d$ of the radial electric field
at the inner edge of SOL ($\left.\frac{\partial v_{y}}{\partial x}\right|_{a}=-\frac{a\varOmega}{d}$)
and the asymptotic value of the electrostatic potential at the outer
edge of SOL, $\varphi_{0}$. Then the matching conditions are 
\[
M_{\nu}=-2\pi a^{3}\frac{\eta}{d}\varOmega,
\]
and, assuming approximately exponential profile of $\phi$ (like in
the recycling regime of SOL), 
\[
\varphi_{*}\approx\varphi_{0}-\frac{Bda}{c}\varOmega.
\]
The effective viscosity, $\eta$, is also difficult to measure. However,
it can be found from the value of $M_{\nu}$ via measurements of $\varOmega(z)$
profiles. This relationship is discussed below.

\begin{figure}
\includegraphics[width=0.9\textwidth]{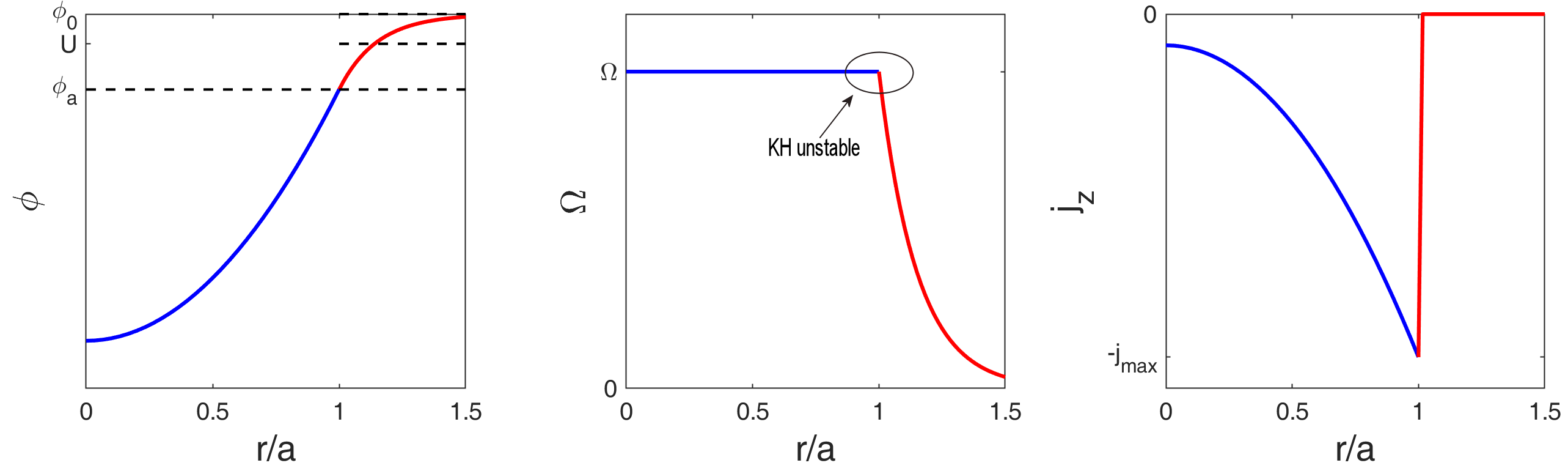}

\caption{\label{fig:Matching-the-discharge}Matching the discharge core ($r<a$)
and the SOL layer.}
\end{figure}

The scheme of matching the rigidly rotating discharge core to the
SOL is shown in Fig.\ref{fig:Matching-the-discharge}. The $\varOmega(r)$
profile in the matching zone $r\approx a$ satisfies the necessary
condition for the Kelvin-Helmholtz instability. Indeed, its smoothed
version contains the inflection point. In SMOLA the core edge appears
to be the source of fluctuations \citep{Tolkachev2024ElectromagneticOA}.
As a result, viscosity and resistivity there are likely anomalous.
As shown in the graph for the axial current density, $j_{z}$, its
absolute value in the edge zone is enhanced, while the axial ohmic
field can cause the axial dependence of $\varphi_{*}(z)=\phi(a,z)$
and hence, $\varOmega(z)$.

Axial dependence of $\varOmega$ can be deduced from the torque balance
equation, $\partial\varPhi(z)/\partial z=M_{A}+M_{\nu}$. Using the
``experimental'' matching conditions, in terms of $\varOmega(z)$
it looks as: 
\begin{equation}
\frac{\pi\sigma B^{2}}{12c^{2}}a^{5}(a+d)\frac{\partial^{2}\varOmega}{\partial z^{2}}-\frac{\partial}{\partial z}\left(G\varOmega\right)-\frac{2\pi a^{3}}{d}\eta\varOmega=\frac{\pi\sigma B}{4c}a^{4}\varphi_{0}''(z).\label{eq:a}
\end{equation}
The physical meaning of the equation terms here is as follows: 
\begin{enumerate}
\item The torque via Ampere's force due to the biasing of end electrodes.
In ideally conducting plasma ($\sigma\rightarrow\infty$) it is dominant
and ensures $\varOmega=const$ in $z$. 
\item The divergence of the axial flow of the angular momentum with the
plasma flow, ensures its continuity. 
\item The frictional torque caused by interaction with the SOL plasma at
the core edge. 
\item The effect of the axially distributed SOL biasing, if present (via
the limiters in the line-tying regime, or via the conducting shell). 
\end{enumerate}
The above equation for $\varOmega(z)$ can be rewritten in the normalized
form as 
\begin{equation}
L^{2}\frac{\partial^{2}\varOmega}{\partial z^{2}}-\varOmega=\frac{\partial}{\partial z}\left(\hat{G}\varOmega\right)+D.\label{eq:Lz}
\end{equation}
Here $L$ is the characteristic axial variation length, 
\[
L=c_{A}\sqrt{\tau_{\sigma}\tau_{\eta}/24}\propto Ba^{3/2}\sigma^{1/2}\eta^{-1/4},
\]
where $c_{A}$ is the Alfven velocity, $\tau_{\sigma}$ is the resistive
time, and $\tau_{\eta}$ is the viscous momentum transport time: 
\[
\tau_{\sigma}=\frac{4\pi\sigma}{c^{2}}a^{2},\quad\tau_{\eta}=\frac{\rho d(a+d)}{\eta},
\]
\[
\hat{G}=\int_{0}^{1}v_{z}\tau_{\eta}\left(\frac{\rho r^{3}}{\rho_{*}a^{3}}\right)\mathrm{d}\left(\frac{r}{a}\right)
\]
is the normalized axial plasma flux, and 
\[
D=\frac{\sigma Bad}{8c\eta}\varphi_{0}''(z).
\]

Eq.(\ref{eq:Lz}) can be solved with suitable boundary conditions
(modeling the cathode and the end plates) along the discharge, if
its coefficients are defined. Unfortunately, they are not, as it is
very difficult to measure or calculate the anomalous plasma viscosity.
In these circumstances it would be more effective to use Eq.(\ref{eq:Lz})
to find $L$ and then $\eta$ via measurements of $\varOmega(z)$.

The axial distribution of the electric field depends on the plasma
resistivity as well. It is easier to measure than the viscosity, but
there may exist another complication. Namely, the Ohm's law for discharges
with emissive cathodes (like in SMOLA) may need modification to include
the electron inertia. Indeed, the parabolic form of the plasma potential
in radius suggests that the sheath potential at the cathode edge in
radius may be large, of the order of the full biasing potential. Then
at least part of the axial electron current may be sustained by the
beam effect, which is not included in the present model.

\section*{Conclusion }

We presented two theoretical models of the plasma rotation induced
by biasing in axially symmetric traps. The difference between these
models is in the role of the effective plasma viscosity that describes
the radial transport of the angular momentum.

For fusion-relevant plasma parameters the rotation driven by biasing
or by external sources of momentum can be described as axially rigid.
It is superimposed by the non-rigid rotation due to the ambipolar
plasma potential, defined by the equilibrium plasma distribution.
The radial structure of the driven rotation is complicated, its temporal
evolution depends on the biasing geometry as well as the transport
of the angular momentum.

In low-temperature or small-radius discharges (such as in SMOLA) the
viscosity can cause the plasma rotation to be radially rigid, while
in presence of the finite parallel conductivity the rotation frequency
may vary along the device. In this model the loss of the angular momentum
is due to friction with the SOL plasma at the edge. The axial decay
length of the rotation frequency is $L=c_{A}\sqrt{\tau_{\sigma}\tau_{\eta}/24}$.
The area of contact of the core plasma with the SOL may be turbulent
(this increases the effective viscosity) due to the Kelvin-Helmholtz
instability. Faster rotation of the core can be achieved by increasing
the magnetic field, by lowering the effective viscosity and by suppressing
the momentum loss rate in SOL. The latter requires lowering the recycling
rate and using limiter designs that prohibit radial currents to reduce
the line-tying effects.

Biasing the plasma-facing electrodes remains the simplest and cost-effective
means to drive the plasma rotation in linear traps. It is less obvious
that improvement of the plasma parameters makes it more rather than
less efficient since the necessary biasing currents drop. Estimates
by \citet{Skovorodin2023GasDynamicMT} show that the biasing-based
``vortex confinement'' can be realized in the next-generation fusion
trap, GDMT.

\section*{Acknowledgments}
The author is grateful to the experimental groups of GDT and SMOLA traps in the Budker INP, especially to 
A. Sudnikov, A. Inzhevatkina and E. Soldatkina for providing and discussing the relevant experimental data.

This work was supported by Russian Science Foundation Grant no. 22-12-00133

%\begin{thebibliography}
\medskip{}

 \bibliographystyle{apalike}
\bibliography{ccc}

%\end{thebibliography}
\end{document}